\documentclass[namedreferences]{SolarPhysics}
\usepackage[optionalrh]{spr-sola-addons} 
\usepackage{epsfig}          
\usepackage{graphicx}        
\usepackage{color}           
\usepackage{url}             
\definecolor{orange}{rgb}{1,0.5,0}

\newcommand{\etal}{{\it et al.}}

\renewcommand{\vec}[1]{ {\mathbf #1} }

\newcommand{\adv}{{\it Adv. Space. Res.}}

\newcommand{\aap}{{\it Astron. Astrophys.}}

\newcommand{\apj}{{\it Astrophys. J.}}
\newcommand{\apjl}{{\it Astrophys. J. Lett.}}

\newcommand{\grl}{{\it Geophys. Res. Lett.}}

\newcommand{\jgr}{{\it J. Geophys. Res.}}

\newcommand{\nat}{{\it Nature}}

\newcommand{\solphys}{{\it Solar Phys.}}

\begin{document}

\begin{article}

\begin{opening}

\title{Homologous Flares and Magnetic Field Topology in Active Region NOAA 10501 on 20 November 2003}

\author{R.~\surname{Chandra}\sep
        B.~\surname{Schmieder}\sep C.H.~\surname{Mandrini}\sep P.~\surname{D\'emoulin}
\sep E.~\surname{Pariat} \sep T.~\surname{T\"{o}r\"{o}k}\sep W.~\surname{Uddin}}

\runningauthor{Chandra et al.}
\runningtitle{Homologous Flares and Magnetic field Topology}

\institute{R. Chandra \sep B. Schmieder \sep E. Pariat \sep P. D\'emoulin \sep T. T\"{o}r\"{o}k$^*$\\
Observatoire de Paris, LESIA, UMR8109 (CNRS), F-92195, Meudon Principal
Cedex, France\\
$^*$Current address: Predictive Science, Inc., 9990 Mesa Rim Road,
San Diego, CA 92121, USA \\
email: \url{chandra.ramesh@obspm.fr}\\
\medskip
C.H. Mandrini\\
Instituto de Astronom\'\i a y F\'\i sica del Espacio (IAFE), CONICET-UBA, Buenos
Aires, Argentina \\
\medskip
W. Uddin\\
Aryabhatta Research Institute of Observational Sciences (ARIES), Nainital - 263 129, India
              \\
             }

\begin{abstract}
We present and interpret observations of two morphologically
homologous flares that occurred in active region (AR) NOAA 10501 on
20 November 2003. Both flares displayed four homologous H$\alpha$
ribbons and were both accompanied by coronal mass ejections (CMEs).
The central flare ribbons were located at the site of an emerging
bipole in the center of the active region.  The negative polarity of
this bipole fragmented in two main pieces, one rotating around the
positive polarity by $\approx 110^{\circ}$ within 32 hours. We model
the coronal magnetic field and compute its topology, using as
boundary condition the magnetogram closest in time to each flare. In
particular, we calculate the location of quasi-separatrix layers
(QSLs) in order to understand the connectivity between the flare
ribbons. Though several polarities were present in AR 10501, the
global magnetic field topology corresponds to a quadrupolar magnetic
field distribution without magnetic null points. For both flares,
the photospheric traces of QSLs are similar and match well the
locations of the four H$\alpha$ ribbons. This globally unchanged
topology and the continuous shearing by the rotating bipole are two
key factors responsible for the flare homology. However, our
analyses also indicate that different magnetic connectivity domains
of the quadrupolar configuration become unstable during each flare,
so that magnetic reconnection proceeds differently in both events.
\end{abstract}
\end{opening}

\section{Introduction}
\label{S-Introduction}

Solar flares involve a sudden energy release in a restricted volume
of the solar corona, occurring on time scales of minutes up to a few
hours.  Confined flares take place in loops in the lower corona and
the emission largely comes from the plasma in those loops. Eruptive
flares, on the other hand, are typically associated with a coronal
mass ejection (CME) and often occur in conjunction with a
filament eruption. It is now widely accepted that these three
phenomena (eruptive flare, CME, and filament eruption) are different
observational manifestations of a sudden and violent disruption of
the coronal magnetic field, often simply referred to as ``solar
eruption'' (see, {\em e.g.}, \opencite{Forbes00}). Whether or not
all three phenomena occur together appears to depend mainly on
details of the pre-eruptive configuration. Cases in which the
eruption does not evolve into a CME are called ``failed eruption'' (see, {\em
e.g.}, \opencite{Torok05}).

The so-called CSHKP model for eruptive flares was put forward by
\inlinecite{Carmichael64}, \inlinecite{Sturrock66},
\inlinecite{Hirayama74}, and \inlinecite{Kopp76} and has been
refined by many authors since then. This model provides explanations
for all main observational flare features, as for example the
formation and separation of ribbons. However, it does not address
the detailed mechanisms by which eruptions are triggered and driven,
which still belong to the most lively discussed problems in solar
physics. Accordingly, a large variety of theoretical models for the
initiation of eruptions has been proposed in the last decades (for a
recent comprehensive review, see the Introduction in
\inlinecite{Aulanier10}).   Among these models are the
``tether-cutting model'' (hereafter TC; see, {\em e.g.},
\opencite{Moore92}; \opencite{Moore01}) and the ``magnetic breakout
model'' (hereafter MB; see, {\em e.g.}, \opencite{Antiochos98};
\opencite{Antiochos99}). In the following, we briefly describe these
two models since they appear to be consistent with the observations
presented in this paper.

In the TC model, the eruption starts by the interaction of two
low-lying sheared arcades located in a bipolar area in an AR. If
these arcades come into contact, for example due to photospheric
motions, they start to reconnect and progressively form a magnetic
flux rope. After the stabilizing influence of the magnetic field
overlying the newly formed flux rope has been sufficiently weakened
by the reconnection below it (hence the term ``tether cutting''),
the flux rope erupts, evolving into a CME or into a failed eruption
(see Figure 1 in \opencite{Moore01}). Although formulated for a
bipolar configuration, the model works also if the bipole is part of
a larger multipolar active region \cite{Moore06}.

In the MB model, the eruption is initiated at a magnetic null point (or
null line) located above a sheared magnetic arcade in a quadrupolar
configuration. If the arcade starts to rise, for example due to further
shearing by photospheric motions, its expansion triggers current sheet
formation and reconnection at the null point. The reconnection
progressively decreases the magnetic tension of the overlying magnetic
field, allowing the arcade to erupt. The expansion of the arcade leads
to the formation of a flux rope that erupts as part of the arcade.

We note that both models do not directly address the physical
mechanism by which the flux rope eruption is driven. This mechanism
is most likely a ``catastrophic loss of equilibrium'' or
the ideal MHD (``torus'') instability related to it ({\em e.g.,}
\opencite{for91}; \opencite{kli06}; \opencite{Aulanier10}).

Both the TC and MB models predict a slow reconnection process in the
initial phase of the eruption and a fast one in its main phase. The
latter takes place below the erupting flux and yields two main
ribbons, in agreement with the CSHKP model. The initial
reconnection, however, occurs below the erupting flux in a bipolar
configuration in the TC model and above it in a quadrupolar
configuration in the MB model. As a consequence, the expected
pattern of brightenings in the pre-flare phase is different. Both
models predict the simultaneous appearance of four brightenings in
this phase. In the TC model, these are located within a
bipolar area, pairwise at each side of the polarity
inversion line (PIL, see Figure 1 in \opencite{Moore01}). In the MB
model, they are located at alternating polarities, within a
quadrupolar area of the photospheric magnetic field.

To set constrains on the energy release mechanism and initiation
process during flares, it is helpful to combine the analysis of the
morphology and temporal evolution of the event with the computation
of the magnetic topology of the AR where it occurs ({\em e.g.},
\opencite{Demoulin07}; \opencite{Mandrini10}). Active region
magnetic fields can contain separatrices and, more generally,
quasi-separatrix layers (QSLs). Separatrices are locations where the
connectivity of magnetic field lines is discontinuous. They are
formed by field lines that either pass through a null point or
through a bald-patch neutral line (defined by field lines tangent to the
photosphere at a PIL). QSLs are a generalization of separatrices;
they are defined as regions where the connectivity of magnetic field
lines changes drastically, but can be continuous
(\opencite{Priest95}; \opencite{Demoulin96a};
\opencite{Demoulin96}). The relationship between separatrices/QSLs
in 3D and observations has been explored for a large variety of
solar magnetic configurations. H$\alpha$ and UV flare ribbons have
been found along, or next to, the photospheric or chromospheric
traces of QSLs \cite{Demoulin97,Mandrini97,Bagala00,Masson09}. In
addition, electric current concentrations were found along the
photospheric traces of QSLs \cite{Demoulin97,Mandrini97}. The
release of free magnetic energy, associated with these currents, can
occur at QSLs when their thickness, and that of their associated
current layers, is small enough for reconnection to take place, as
recently shown by MHD simulations of flare-like magnetic
configurations (\opencite{Aulanier05}; \opencite{Aulanier06b}).

Occasionally flares and CMEs can occur in the same AR within minutes
or hours. If such events exhibit a similar morphology, for example
similar locations and shapes of flare brightenings, they are called
homologous. This repetitive character of flares was first discussed
by \inlinecite{Waldmeier38}. Definitions of homologous flaring
include similarities in H$\alpha$ or EUV brightenings, in soft X-ray
structures, and/or radio bursts with similar dynamic spectra ({\em
e.g.,} \opencite{Woodgate84}; \opencite{Machado85};
\opencite{Martres89}). Homologous flares are particularly
interesting phenomena since they pose some challenging questions
about the conditions that lead to flaring: do they occur since not
all of the available free magnetic energy is released or because
free energy is continuously supplied to the AR (see, {\em e.g.},
\opencite{Bleybel02}; \opencite{Schrijver09})?
A quantitative study of homologous flares is presented by \inlinecite{takasaki04}.
In particular, they found that the separation velocity of the flare ribbons has a
good correlation with the soft X-ray light curve, and they derived a similar coronal magnetic 
field strength for three homologous flares, a result which support magnetic reconnection for 
the energy release mechanism.
Moreover, recent numerical simulations have demonstrated that eruptions can indeed occur
repeatedly if flux continues to emerge into the corona
\cite{MacTaggart09}, or if the coronal magnetic field is
continuously sheared by photospheric motions \cite{DeVore08}.

On the other hand, the fact that homologous flares are morphologically similar is an
indication of the key role played by the magnetic field
topology (including QSLs), since it is mainly determined by the global distribution
of the photospheric polarities (see examples in \opencite{Machado88};
\opencite{Mandrini91}, \citeyear{Mandrini97}).
However, though morphologically similar, some homologous flares can
produce different soft X-ray light curves, implying differences in the
impulsiveness of energy release (\opencite{Ranns00}; \opencite{Sterling01})

In this paper we study observations of two homologous flares that
occurred in AR NOAA 10501 on 20 November 2003. A multi-wavelength
study of these events, including the associated filament evolution
and CMEs, and an analysis of a magnetic cloud associated with the
second event, was recently presented by \inlinecite{Kumar09}. Here
we focus on the flare evolution, morphology, and homology, and we
compare our interpretation of the two events with the interpretation
by \inlinecite{Kumar09}. While these authors attributed the flare
trigger for both events to the dynamic interaction of two filaments,
we rather believe that an emerging and rotating bipole in the center
of the AR was at the origin of the observed eruptive activity.

A summary of the events and a description of the analyzed data are presented
in Section~\ref{obs}.   The observational details of both flares and the
evolution of the photospheric magnetic field
are described in Section~\ref{flares}. Section~\ref{topo} presents the
magnetic field model and topology of the AR, together with a possible
explanation for the evolution and morphology of the flares. Finally,
we present our conclusions in Section~\ref{conclusion}.

\begin{figure} 
\centerline{\hspace*{-0.05\textwidth} $\color{black}
\bf\put(275,200){\vector(-1,-1){20}}$ $\color{black}
\bf\put(130,180){\vector(-1,-1){20}}$
\includegraphics[width=0.9\textwidth,clip=]{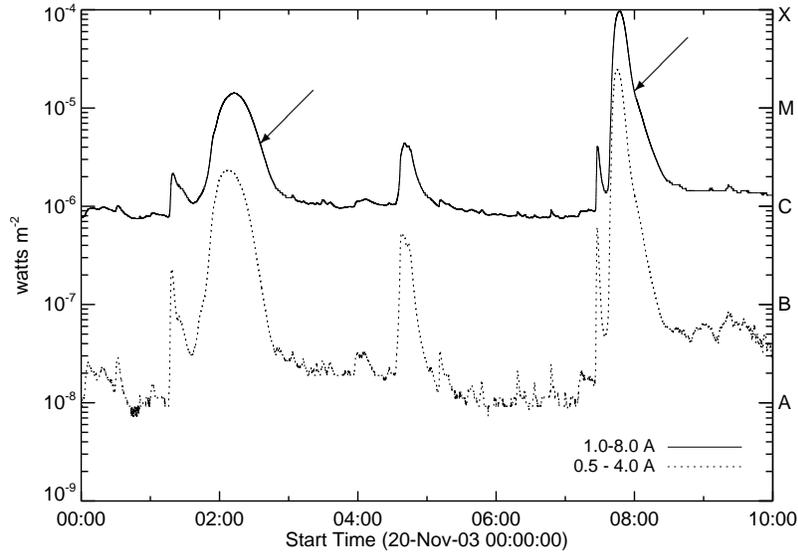}}
\caption{Temporal evolution of the solar soft X-ray flux as observed
by GOES 12 in the 0.5-4 \AA\ and 1-8 \AA\ bandwidths on 20 November
2003. The two studied flares, occurring in AR 10501, are pointed at
by arrows. The M1.4 event occurred at 02:12 UT and the M9.6 at 07:45
UT. The C4.3 event at $\approx$ 04:32 UT occurred in AR 10508.}
\label{goes}
\end{figure}

\section{Summary of the Activity in AR 10501 and Observational Data}
\label{obs}
%
\subsection{Activity in AR 10501 on 20 November 2003}
\label{summary}

Active region NOAA 10501 was one of the most complex and eruptive
regions during the decay phase of solar cycle 23. This AR was
the successor of the also very flare productive active region NOAA 10484, as
it was named during the previous solar rotation. AR 10501 produced 12 M-class flares,
some of them accompanied by CMEs, from 18 to 20 November 2003. In
particular, the early events on 18 November 2003 have been
extensively studied (\opencite{gopalswamy05};
\opencite{Yurchyshyn05}; \opencite{Mostl08};
\opencite{Srivastava09}; \opencite{Chandra10}). The peculiarity of
this AR on that day was that it contained regions of
different magnetic helicity signs, as was discussed by \inlinecite{Chandra10}.
This characteristic seems to be present also on 20 November 2003
(see Section~\ref{topo}).

On 20 November 2003, two homologous flares took place within
approximately five hours in the region, which was located around N03
W05 at that time (see Section~\ref{flares}). The flares were,
respectively, classified as M1.4 and M9.6 from the GOES soft X-ray
emission (Figure~\ref{goes}). The first flare (M1.4) was preceded by
a pre-flare phase starting at $\approx$ 01:30 UT. Its main phase
started at 01:45 UT with a relatively gradual onset, peaked at 02:12
UT, and ended around 02:40 UT. The flare was accompanied by a very
dynamic behaviour of the AR filaments and by a CME (see
Section~\ref{filaments} for details).

The second flare was preceded by a weak (C3.8) precursor starting
at around 7:25 UT. The flare main phase began with an impulsive onset at 07:35 UT,
peaked at 07:45 UT, and reached class M9.6. Finally, the
flare ended gradually at around 08:40 UT. This flare was accompanied by
successive filament eruptions and by a CME (see
Section~\ref{filaments2} for details.)

Though the flares were morphologically similar, the soft X-ray GOES
light curve indicates a more gradual energy release during the first
flare than during the second one. Another flare (of C-class) started
at $\approx$~4:32 UT (Figure~\ref{goes}). However, it was located in
another region (AR 10508). Just after this flare, around 5:12 UT, a
small flare was occuring in AR 10501 (see the small peak in
Figure~\ref{goes}). This flare took place at the same location, and
displayed similar ribbons, as the M1.4 and M9.6 flares (see the EIT
195 \AA\ observations at
\url{http://www.ias.u-psud.fr/eit/movies/}). Around 23:47 UT,
another homologous flare occurred, indicating that the magnetic
configuration of AR 10501 was still similar. We do not analyze these
two flares because of the lack of H$\alpha$ data and because of the
low cadence of the EIT data (at best 12 min in 195 \AA ).

\subsection{The Analyzed Data}

For our study, we mainly use H$\alpha$ data from the Aryabhatta
Research Institute of Observational Sciences (ARIES) in Nainital,
India. These data were obtained by a 15-cm f/15 coud\'e telescope
(pixel size 1$''$) and a Lyot filter centered at the H$\alpha$ line \cite{Uddin04}.
The cadence of the H$\alpha$ images varies from 30 s to 5 min.
Furthermore, we use data from the Solar and Heliospheric
Observatory/Michelson Doppler Imager (SOHO/MDI;
\opencite{Scherrer95}) with a pixel size of 1.98$''$ and a cadence
of 96 minutes. The H$\alpha$ data were co-aligned with MDI
magnetograms by comparing them with white-light images from
SOHO/MDI.

We complement our analysis with reconstructed images from the Reuven
Ramaty High-Energy Solar Spectroscopic Imager (RHESSI;
\opencite{Lin02}), which partly observed the first flare. We
reconstructed the images in the 20-60 keV energy band from seven
collimators (3F to 9F) using the {\it CLEAN} algorithm, which yields
a spatial resolution of $\approx 7''$ \cite{Hurford02}.

\begin{figure} 
\centerline{
\includegraphics[bb=142 184 463 664,clip=]{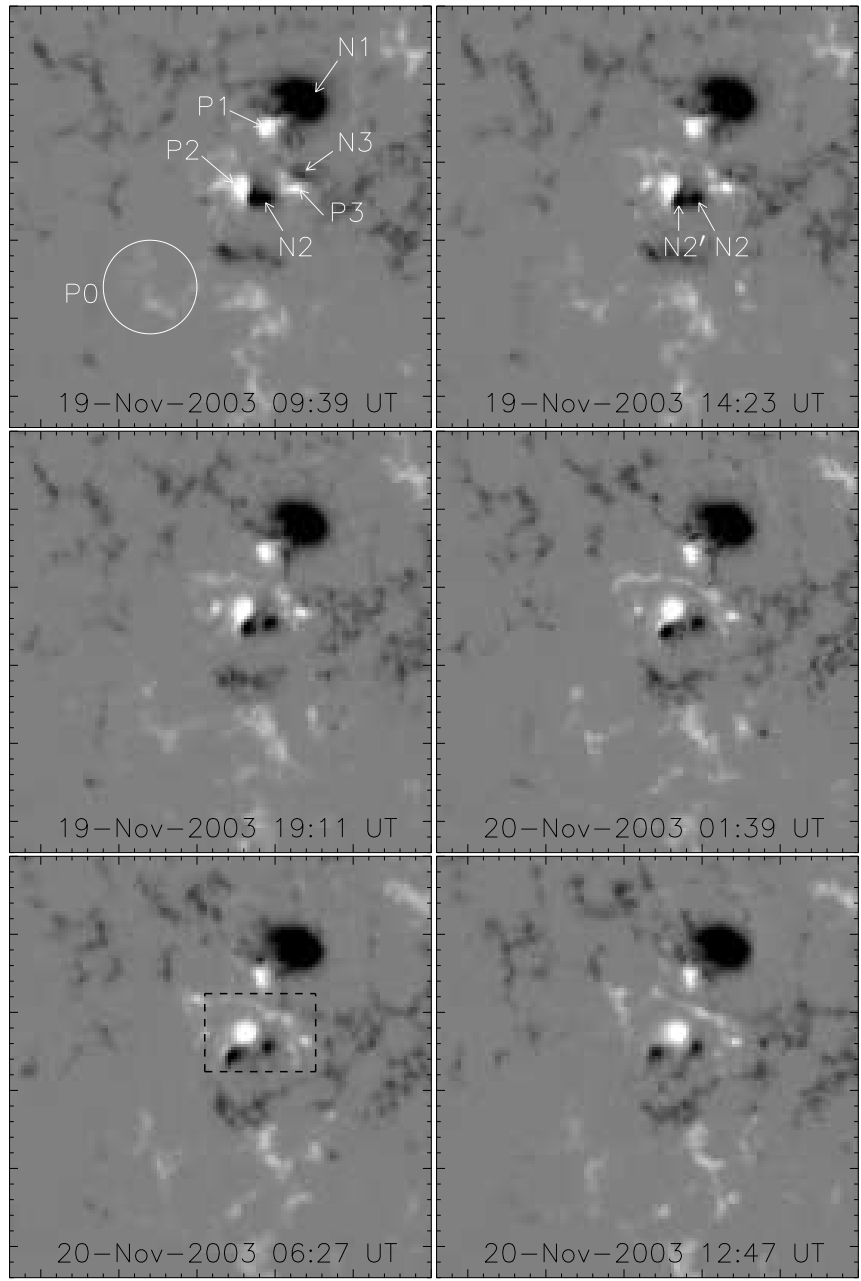}
}
\caption{Evolution of the line-of-sight magnetic field on 19 and 20 November 2003. The
dashed box in the bottom left panel shows the region where N$2^\prime$ rotates around
P2. P0 marks the positive polarity network between the two filaments
(see Figure~\ref{flare1}a).
The field of view of the images is the same as in Figure~\ref{preflare}
}
   \label{mdi}
   \end{figure}

\begin{figure}[t] 
\centerline{\hspace*{0.04\textwidth}
               \includegraphics[width=0.90\textwidth,clip=]{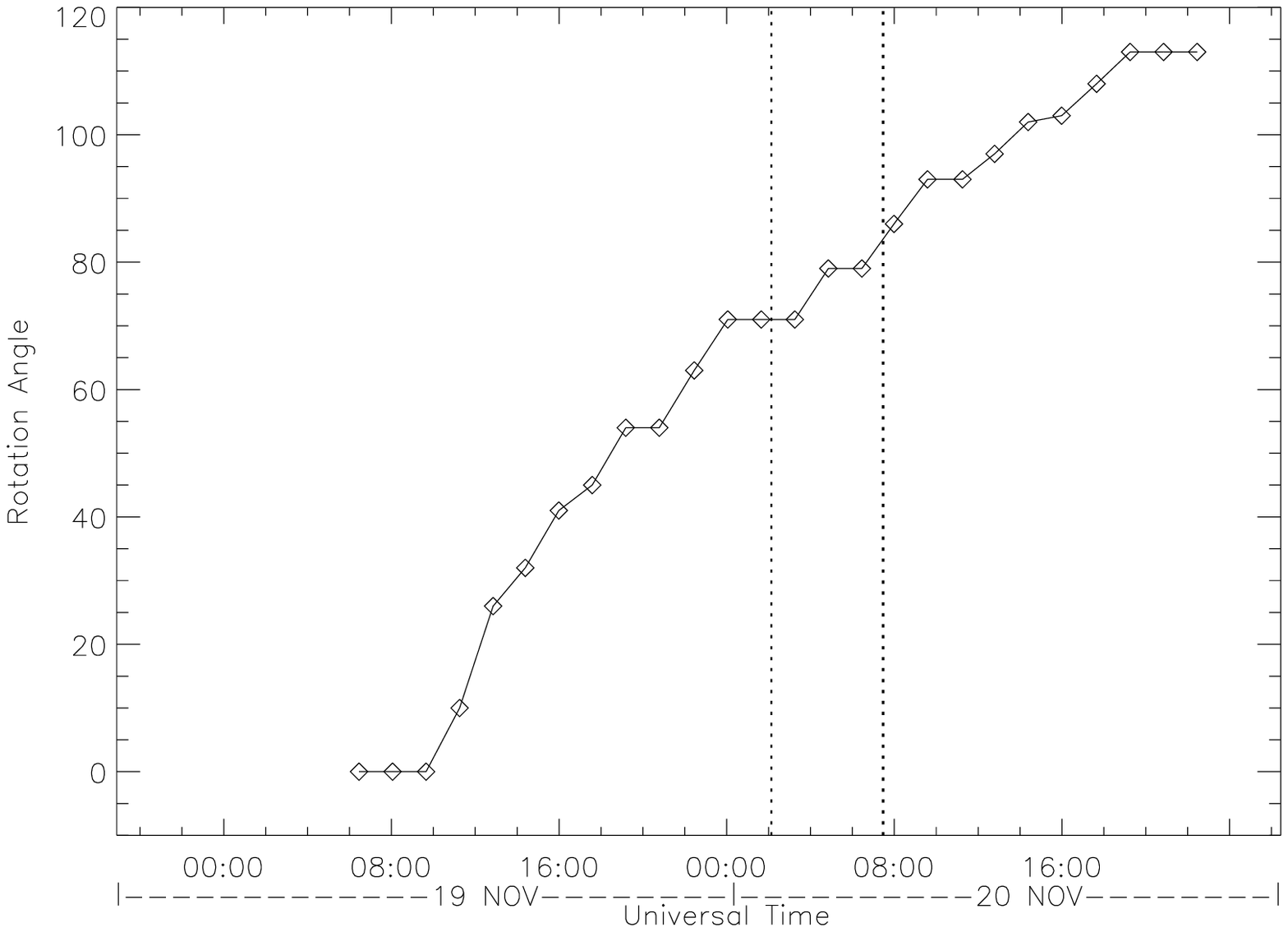}
}
\caption{Temporal evolution of the rotation angle ($\widehat{\rm N2P2N2^\prime}$
in degrees) of N2$^\prime$ around P2 (see Figure~\ref{mdi}). The vertical lines
show the peak times of the first and second flares, respectively. }
   \label{angle}
   \end{figure}

\section{Photospheric Field and Chromospheric Flare Evolution}
\label{flares}

\subsection{Evolution of the Photospheric Magnetic Field} 
\label{mdi_evolution}

Figure~\ref{mdi} presents the evolution of the magnetic field on 19
and 20 November 2003, before, during and after the flares. We have
numbered the six observed main polarities P1, P2, P3 (positive
polarities) and N1, N2, N3 (negative polarities) in the upper left
panel of Figure~\ref{mdi}. We notice a strong diffusion of the
magnetic field around the main negative polarity N1 (which is also
the main spot, see the accompanying MDI movie, ar10501-mdi.mpg). The
bipole N2, P2 in the central part of the active region was already
visible on 15 November, when the region arrived at the eastern limb.
As visible in Figure~\ref{mdi} (and in the MDI movie), the negative
polarity N2 broke into two main pieces, which we named N2 and
N2$^\prime$ in the top right panel. After the breaking, N2 remained
almost at the same location, while N2$^\prime$ rotated around P2.
Many other smaller negative fragments rotated as well, some
preceding, some following N2$^\prime$. By computing the centroid
location of N2, P2, and N2$^\prime$, we measured the rotation angle
$\widehat{\rm N2P2N2^\prime}$. The temporal variation of this angle is
shown in Figure~\ref{angle} from 19 November at 06:27 UT to 20
November at 20:51 UT. The angle started increasing at 09:39 UT on 19
November and it reached $\approx$ 110$^{\circ}$ at 17:39 UT on 20
November, after which the rotation ceased. Both flares studied in
this paper took place during this rotation period of 32 hours.

This continuous rotation is likely to have increased the magnetic field shear along
the magnetic inversion line between the polarities P2 and N2$^\prime$,
and hence the free magnetic energy available for flaring.
This is supported by the Marshall Space Flight Center magnetic
vector data obtained on 19 November at 19:36 UT. The shear map,
shown in Figure 10 of \inlinecite{Kumar09}, indicates that one of
the locations of highest shear angle is along the PIL between N2+N2$^\prime$ and P2. The
clockwise rotation of N2$^\prime$ around P2 implies the injection of negative
magnetic helicity into the configuration.

In addition, one observes a general pattern of dispersion and
cancelation of numerous small negative polarities with positive
polarities of the network at the periphery of the AR in the
south-west (see the MDI movie). This implies that the disperse
positive polarity of the network (P0, see Figure~\ref{mdi})
progressively shrunk both in magnetic flux and surface extension.

\begin{figure} 
\vspace*{-0.05\textwidth}
\centerline{\hspace*{-0.01\textwidth}
\includegraphics[width=0.65\textwidth,clip=]{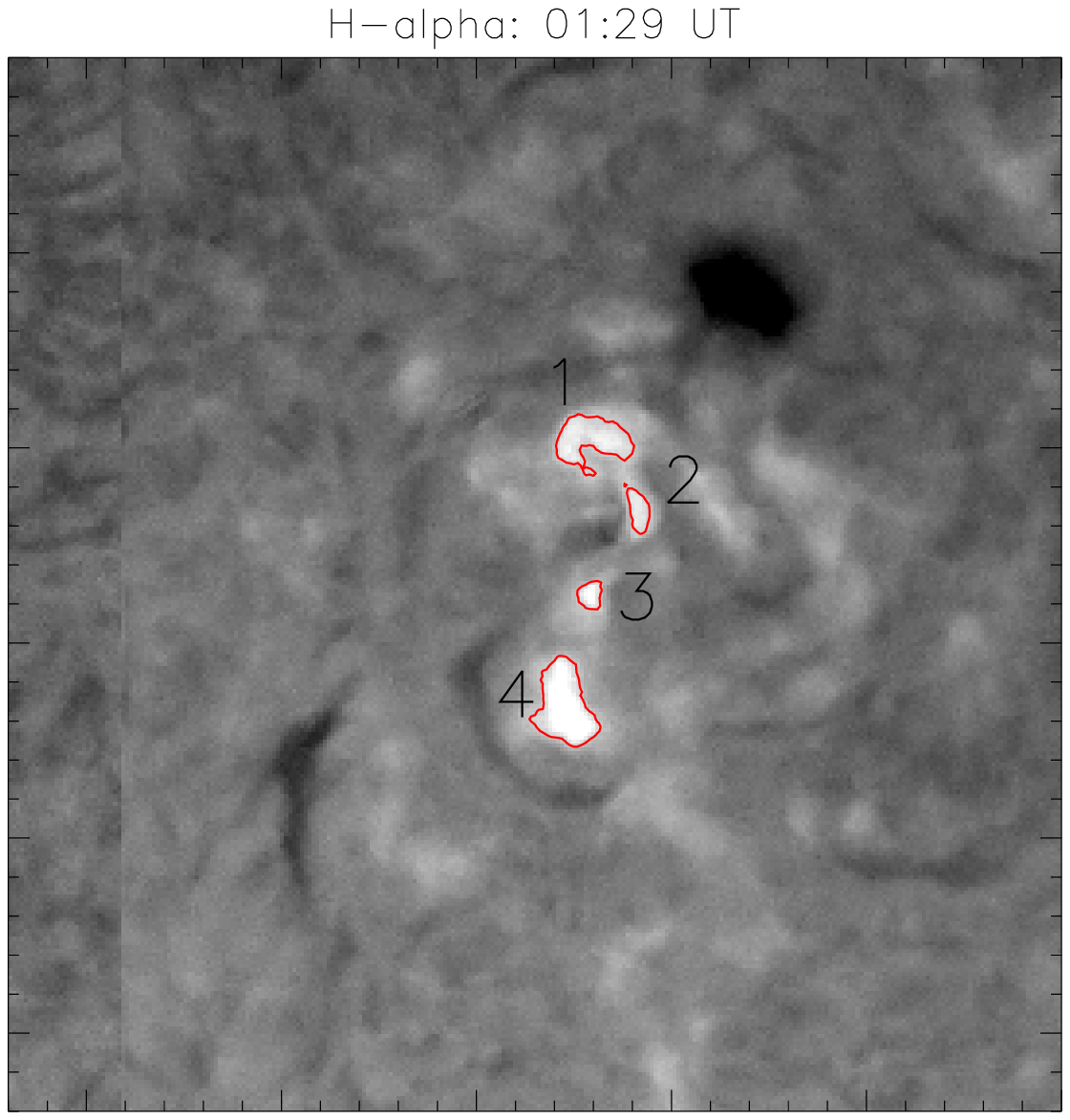}
\color{white} \bf\put(-190,180){(a)}$$
\hspace*{-0.20\textwidth}
\includegraphics[width=0.65\textwidth,clip=]{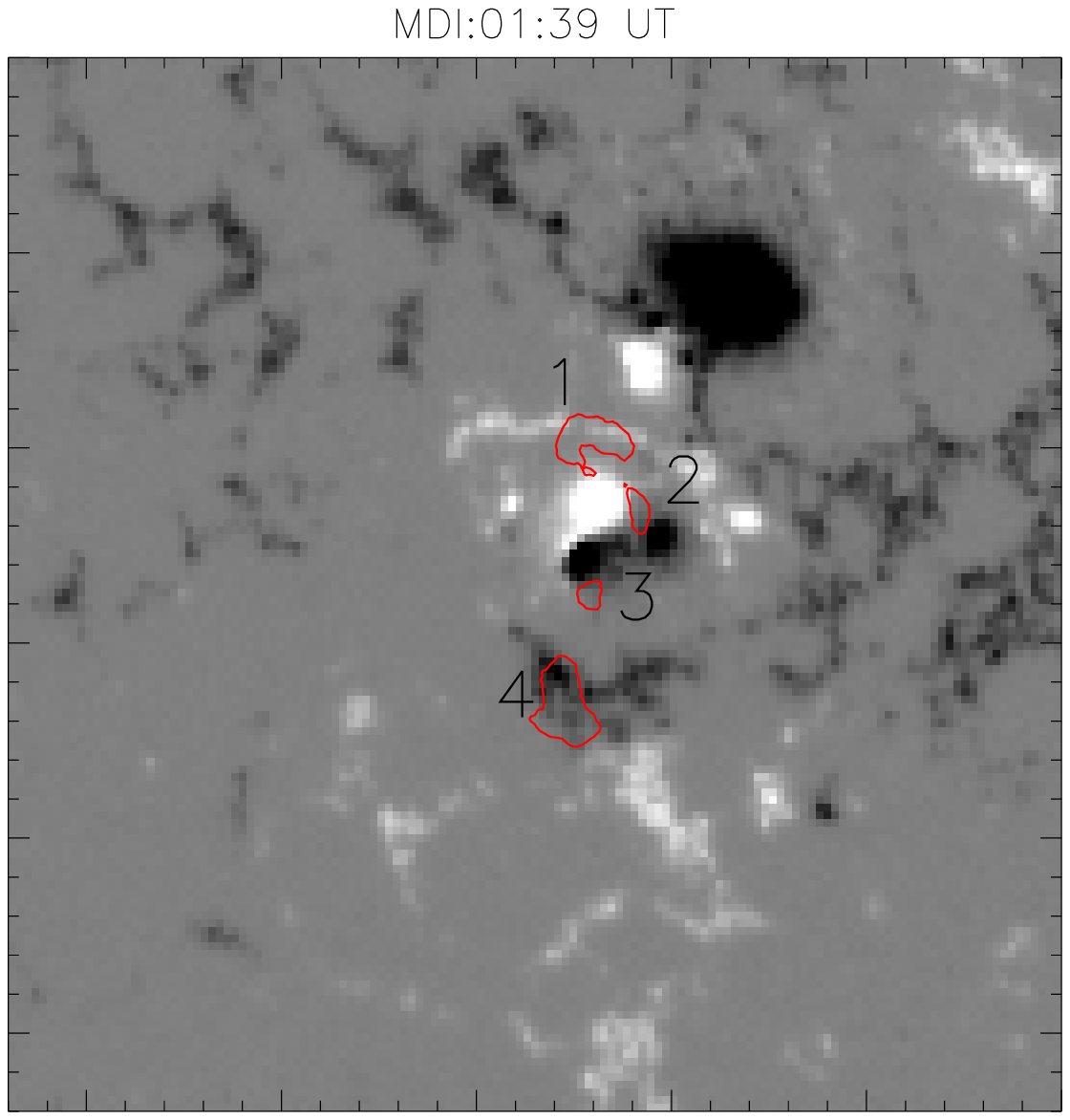}
\color{white} \bf\put(-190,180){(b)}$$ } \vspace*{-0.05\textwidth}
\caption{(a): H$\alpha$ image during the pre-flare phase of the M1.4
event. Four brightenings are numbered as 1-4 and framed in red (by
eye). (b): MDI magnetogram around the same time, overlaid with the
locations of the brightenings. The field of view of the images is
270$''\times$270$''$. } \label{preflare}
\end{figure}

\subsection{The M1.4 Class Flare}
\label{fl1}
%
\subsubsection{Chromospheric Evolution}

During the pre-flare phase of the M1.4 event, at $\approx$ 01:30 UT
(Figure~\ref{goes}), the H$\alpha$ data reveal four small
brightenings, numbered as 1-4 in Figure~\ref{preflare}. Comparing
the locations of these brightenings with the MDI line-of-sight
magnetogram (Figure~\ref{preflare}b), we see that brightenings 1 and
2 are located in positive field regions, whereas 3 and 4 are located
in negative field regions, on opposite sides of the local PIL. The
location of the brightenings, together with their appearance before
the impulsive phase, the photospheric rotation of the negative
polarity N2$^\prime$, and the presence of highly sheared magnetic
field, described in Section~\ref{mdi_evolution}, suggest that the
flare and subsequent CME (see Section~\ref{filaments}) have been
triggered as proposed by the TC model (see
Section\,\ref{S-Introduction}). A similar emission pattern in the
early phase of the X10 flare on 29 October 2003 has been recently
reported by \inlinecite{Xu10}.

\begin{figure} 
\vspace*{-0.50\textwidth}
\centerline{\hspace*{0.08\textwidth}
\includegraphics[width=1.45\textwidth,clip=]{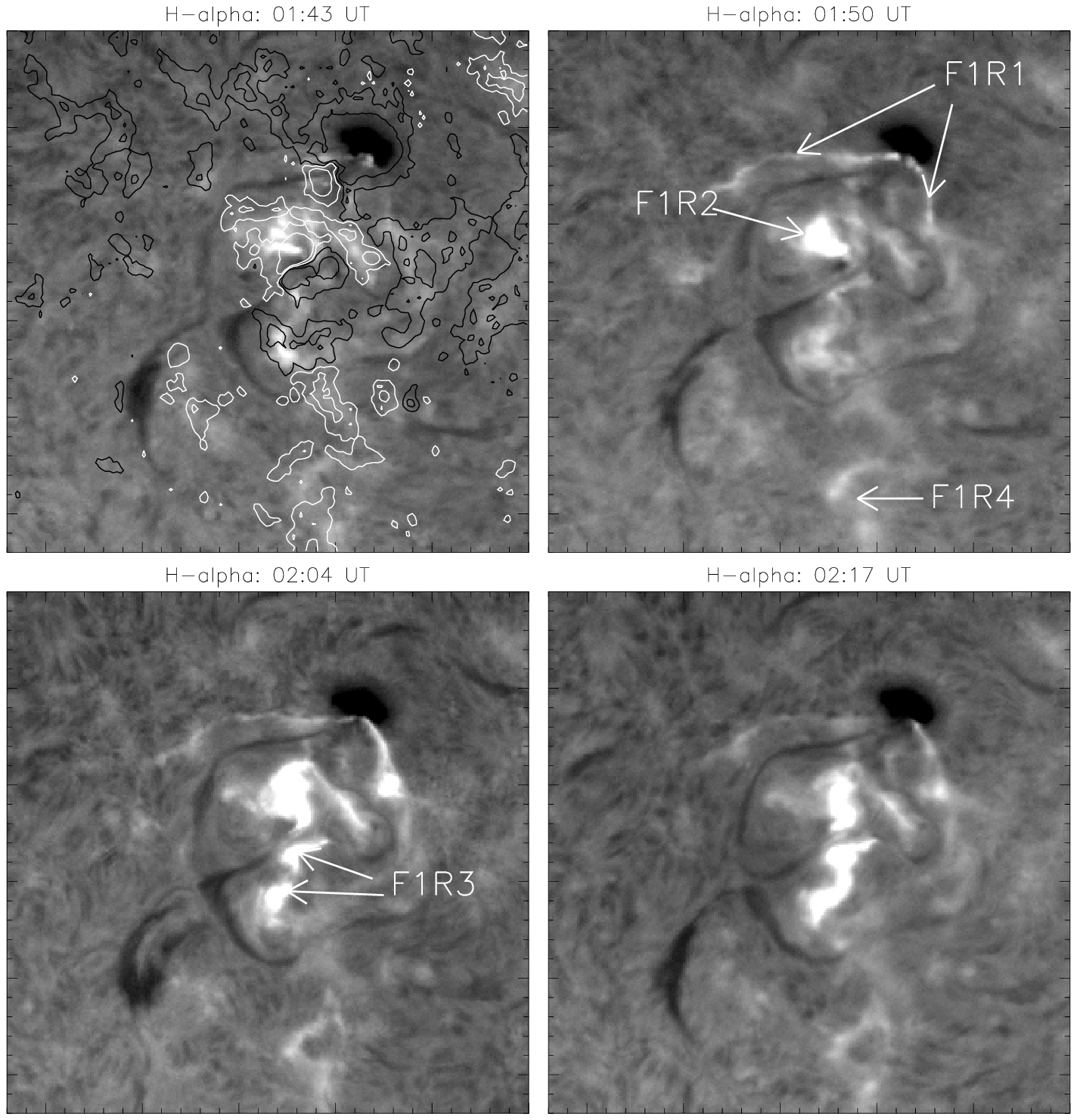}
\color{white} \bf\put(-430,522){(a)}$$ \color{white}
\bf\put(-260,522){(b)}$$ \color{white} \bf\put(-438,342){(c)}$$
\color{white} \bf\put(-268,342){(d)}$$ } \vspace*{-0.5\textwidth}
\caption{H$\alpha$ evolution of the M1.4 flare. The flare ribbons
are marked as F1R1, F1R2, F1R3, and F1R4, respectively. The top left
panel shows overplotted contours of the photospheric line-of-sight
(LOS) magnetic field at 01:39 UT. White (black) contours correspond
to positive (negative) fields. Contour levels $\pm100$ and $\pm500$
G are used. The field of view of the images is the same as in
Figure~\ref{preflare}. } \label{flare1}
\end{figure}

\begin{figure} 
\vspace*{-0.05\textwidth}
\centerline{\hspace*{-0.01\textwidth}
\includegraphics[width=0.7\textwidth,clip=]{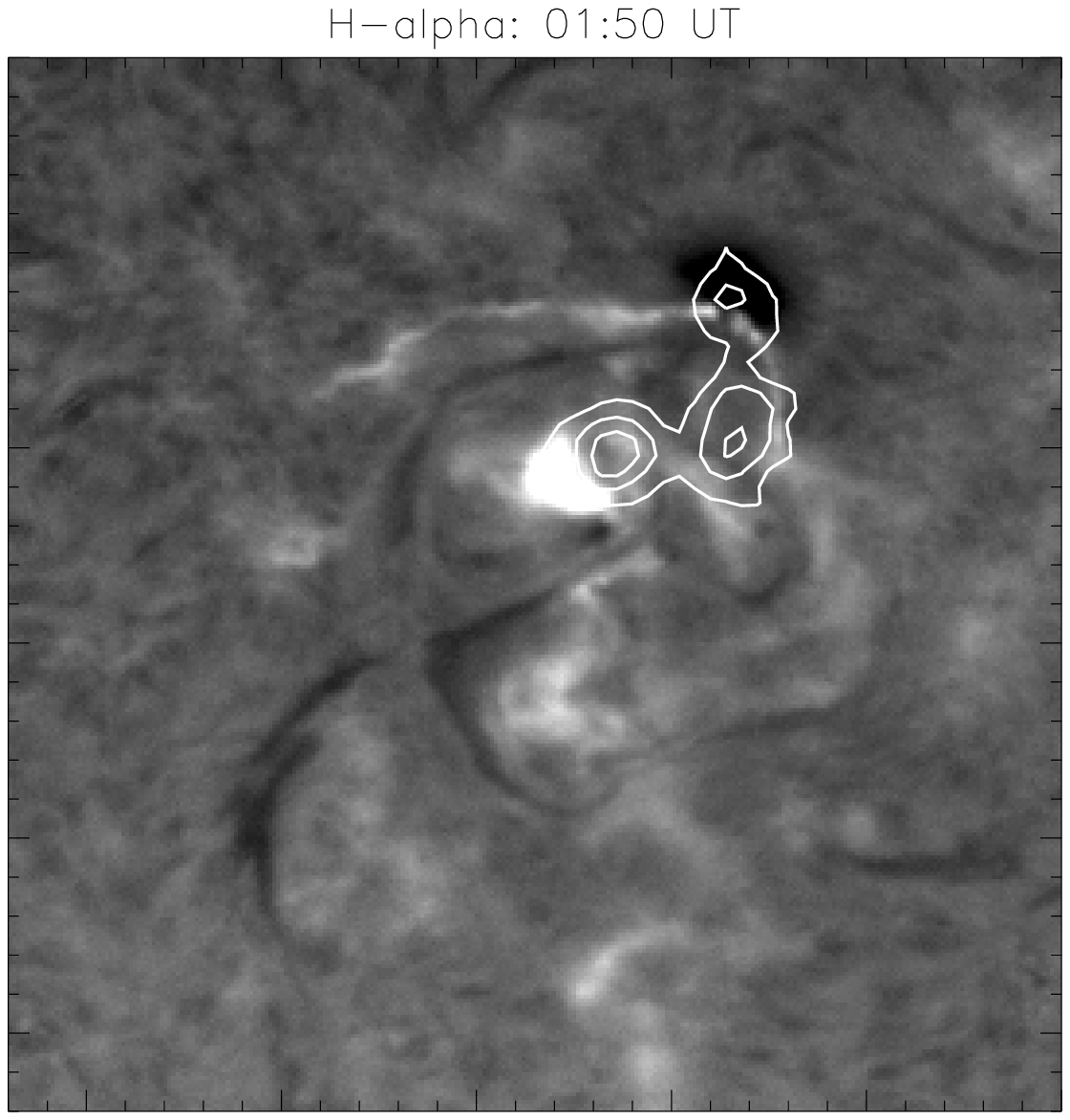}
\color{white} \bf\put(-205,195){(a)}$$
\hspace*{-0.22\textwidth}
\includegraphics[width=0.7\textwidth,clip=]{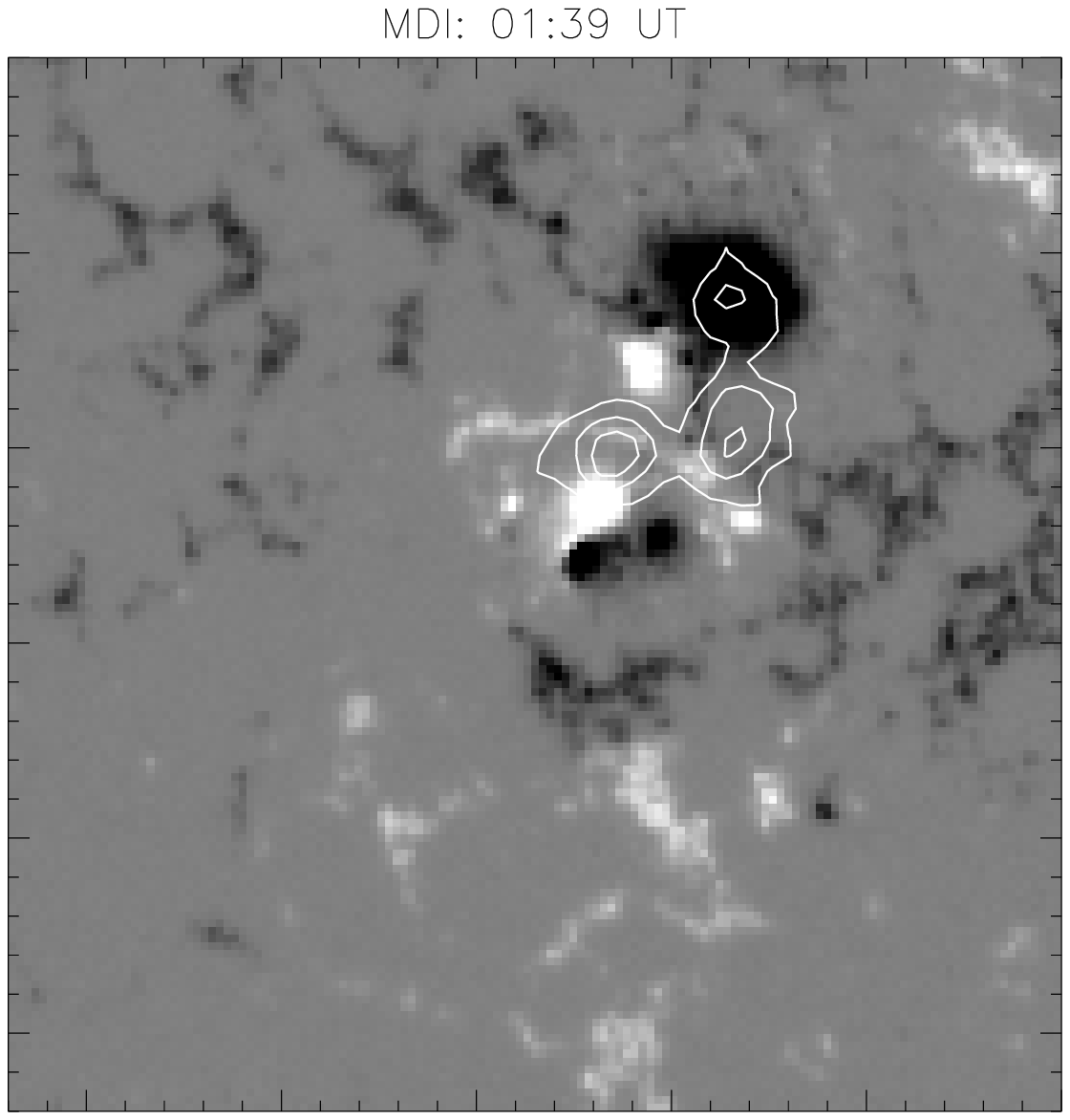}
\color{white} \bf\put(-205,195){(b)}$$ } \vspace*{-0.05\textwidth}
\caption{H$\alpha$ image (a) and MDI magetogram (b),
overlaid by RHESSI hard X-ray (20-60 keV) contours (contour levels
are 40, 60, and 80$\%$ of the peak intensity). The field of view of
the images is the same as in Figure~\ref{preflare}. } \label{rhessi}
\end{figure}

The chromospheric evolution of the flare during its main phase is
shown in Figure~\ref{flare1}. Four distinct ribbons are visible: two
main, inner ribbons F1R2 and F1R3, and two fainter, outer ribbons
F1R1 and F1R4. Comparing Figures \ref{preflare} and \ref{flare1}, it
can be seen that the two main ribbons, F1R2 and F1R3, evolved from
the four brightenings observed in the pre-flare phase, which is
again consistent with the TC initiation model. It can be also seen
that the strong emission set in earlier in F1R2 than in F1R3,
indicating an asymmetry in the magnetic configuration.

The magnetic field of the AR is highly complex, but appears to be
essentially quadrupolar on the large scale (Figure~\ref{preflare}b).
The H$\alpha$ ribbons are located in alternating polarities (along
the north-south direction), as expected for a quadrupolar magnetic
configuration. The appearance of the outer ribbons (F1R1 and F1R4)
suggests that the erupting configuration in between F1R2 and F1R3
has started to reconnect with the overlying magnetic field (see
Section\,\ref{model}).

\begin{figure} 
\vspace*{-0.05\textwidth}
\centerline{\hspace*{-0.01\textwidth}
\includegraphics[width=0.7\textwidth,clip=]{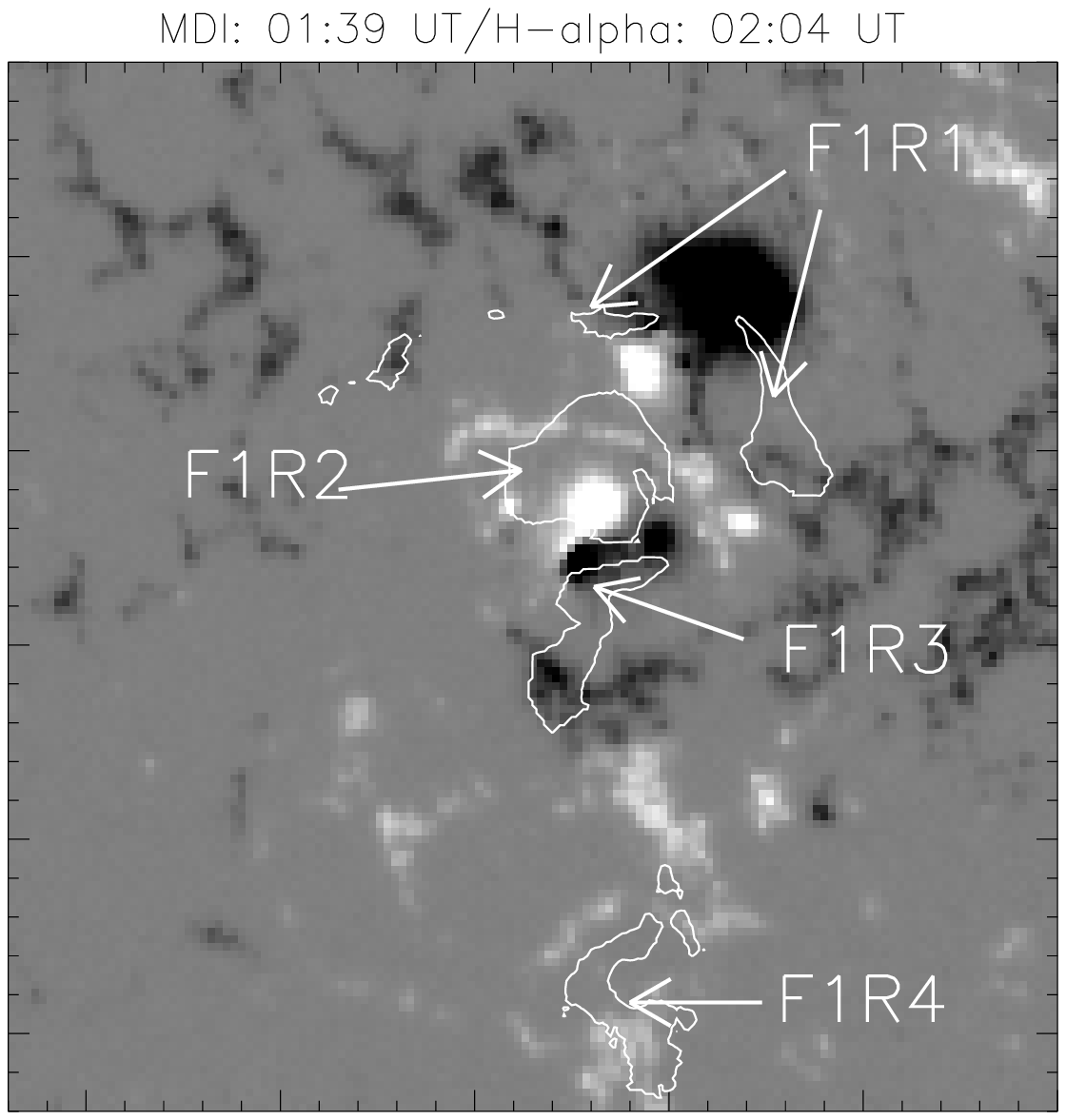}
\color{white} \bf\put(-205,195){(a)}$$
\hspace*{-0.22\textwidth}
\includegraphics[width=0.7\textwidth,clip=]{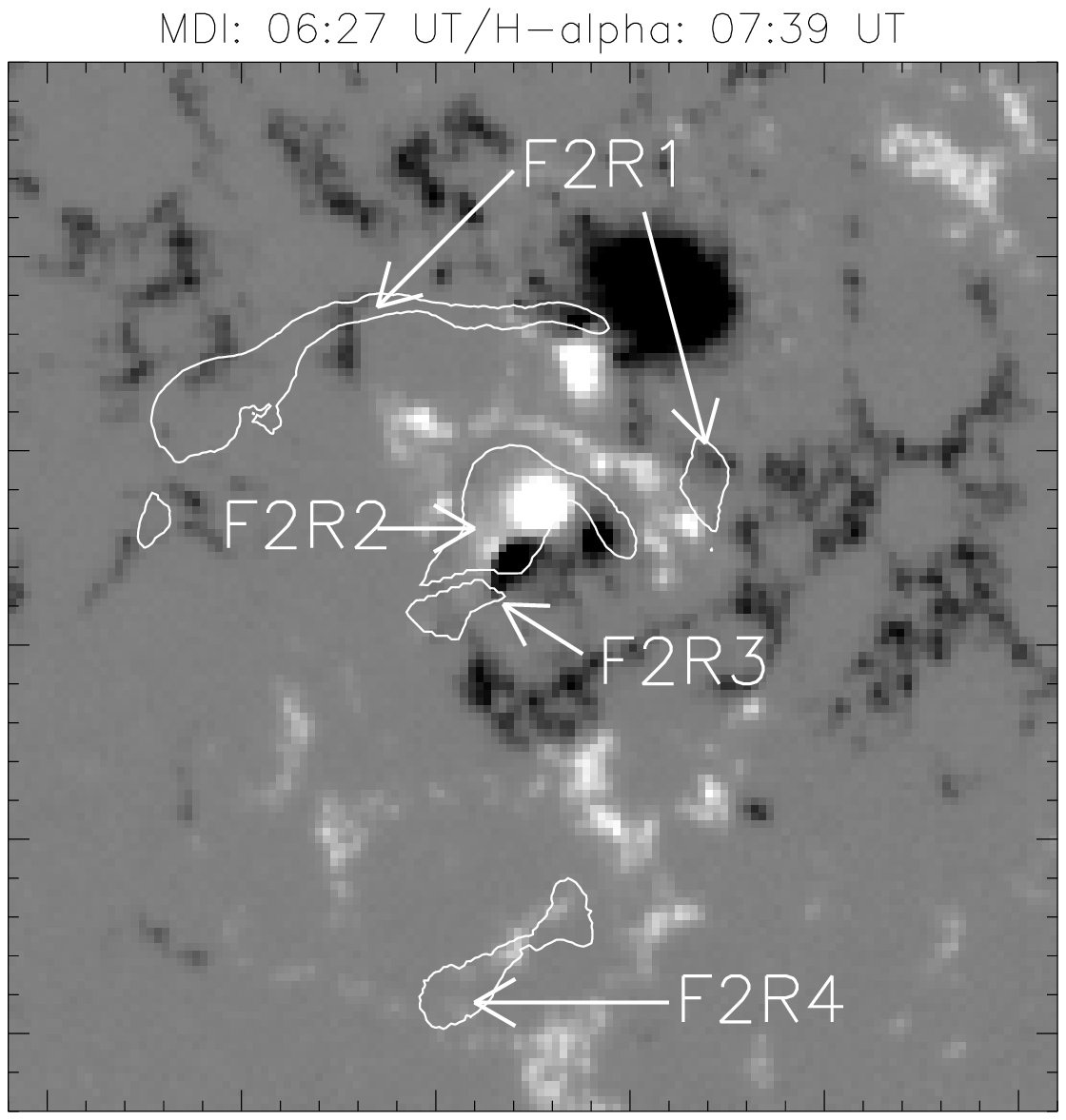}
\color{white} \bf\put(-205,195){(b)}$$
}
\vspace*{-0.05\textwidth}
\caption{Location of the M1.4 (a) and M9.6 (b) flare ribbons
on the closest in time MDI magnetogram. The field of view of the images
is the same as in Figure~\ref{preflare}.}
\label{ribbon_mdi2}
\end{figure}

\subsubsection{Nonthermal Emission}
Around 01:50 UT, during the M1.4 flare peak in soft X-rays
(Figure~\ref{goes}), RHESSI observed three non-thermal sources in
the 20-60 keV energy band (Figure~\ref{rhessi}). The northern and
the eastern sources were located close to the H$\alpha$ ribbons F1R1
and F1R2, respectively, while the third source might have been a
signature of reconnection in the corona, located above newly
reconnected loops connecting F1R1 and F1R2. The magnetic topology
study presented in Section~\ref{model} supports this interpretation.
Unfortunately, we could not produce RHESSI images at other times
during the flare, due to the low count rate.

Comparing the hard X-ray footpoint sources with the photospheric
magnetic field (Figure~\ref{rhessi}b), one can see that the stronger
(eastern) source is located in a region of weak field, whereas the
weaker (northern) source is located in the strong sunspot field.
Such an asymmetry was initially reported by \inlinecite{Sakao94},
who interpreted it as a magnetic mirroring effect. That is, a weaker
field at a footpoint permits a higher electron
precipitation rate than a stronger field.   This suggestion was later
confirmed by several authors (\opencite{Kundu95}; \opencite{Li97};
\opencite{Aschwanden99b}; see also Cristiani {\it et al.} (2008, 2009)
, for similar results in the case of submillimeter observations), and it applies to
this studied event too.

\subsubsection{Filament Dynamic Evolution and CME}
\label{filaments}

As mentioned in Section\,\ref{summary}, the M1.4 flare was accompanied by a CME.
The CME was rather slow ($\approx$ 364 km s$^{-1}$) and had a low
angular width  ($\approx$60$^{\circ}$). It became visible in the
Large Angle and Spectrometric Coronagraph (LASCO,
\opencite{Brueckner95}) C2 field of view at 02:48 UT (see
\url{http://cdaw.gsfc.nasa.gov/CME_list/)}.
Extrapolating back the LASCO height-time plot yields an approximate
start time of the CME at about 01:40 UT, consistent with the time of the
pre-flare phase.

Two elongated filaments were present at the east of the AR. They
displayed a very dynamic behaviour around the time of the flare;
however, they did not participate in the eruption. During the
pre-flare phase, their central sections approached each other; while
during the main flare phase, they seemed to merge, and shortly after
they disconnected again, having different footpoint connections
(Figure\,\ref{flare1}; see also \opencite{Kumar09}).

These two filaments were located above two PILs, which we observed
to shift closer to each other during the rotation of the negative
polarity N2$^\prime$ around P2 (see Section~\ref{mdi_evolution}).
The motion and dispersion of the negative polarities (N2$^\prime$
and smaller ones) is likely to be at the origin of the approach of
the two H$\alpha$ filaments. As the positive magnetic flux between
the filaments (marked P0 in the top left panel of Figure~\ref{mdi})
became too weak, independent magnetic configurations are no longer
possible. As a result, the magnetic configuration reconnects and the
filaments merge to form a new configuration. After the flare, the
two filaments are still present, but display a different pattern: a
$\cup$-shaped filament lies at the north of a $\cap$-shaped one
(Figure~\ref{flare1}d).

In the view of \inlinecite{Kumar09}, the flare was triggered by the
interaction of the two filaments. 
In our view there was no direct cause-and-effect  
between the interaction
of the filaments and the flare, since none of the flare signatures
observed in the available wavelengths was located close to or
magnetically connected to the merging point of the filaments (see
Section~\ref{topo}). The merging of the filaments is expected to
involve magnetic reconnection as proposed by \inlinecite{Kumar09};
however, this reconnection apparently releases much less energy than
the reconnection involved in the flare. Therefore, we believe that
both the dynamic behaviour of the filaments and the flare were
rather independent signatures of a large-scale dynamic
reconfiguration of the AR field, as a result of the continuous
stressing of the field by the emerging and rotating bipole in its
center (see Section\,\ref{mdi_evolution}).  A numerical simulation
of the interaction of these filaments is presented in \inlinecite{torok10}.


\begin{figure} 
\vspace*{-0.50\textwidth}
\centerline{\hspace*{0.1\textwidth}
\includegraphics[width=1.45\textwidth,clip=]{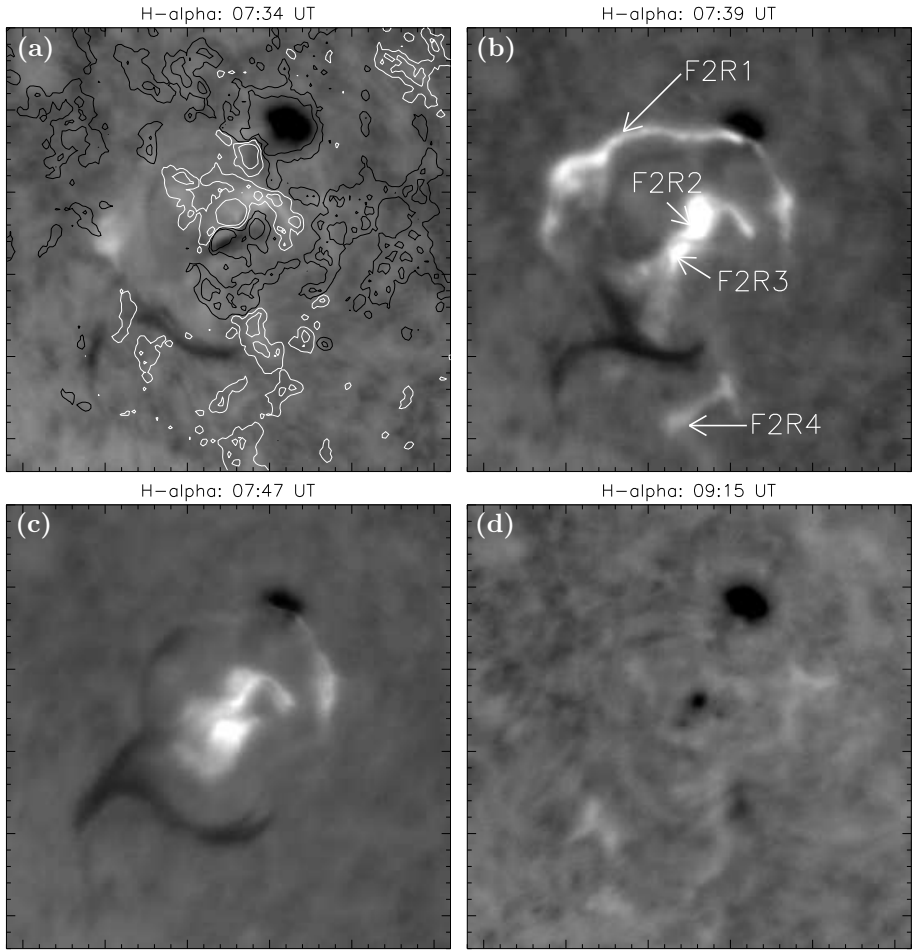}
\color{white} \bf\put(-435,522){(a)}$$ \color{white}
\bf\put(-265,522){(b)}$$ \color{white} \bf\put(-443,342){(c)}$$
\color{white} \bf\put(-273,342){(d)}$$ } \vspace*{-0.5\textwidth}
\caption{Evolution of  M9.6 flare in H$\alpha$. The flare ribbons
are marked by F2R1, F2R2, F2R3 and F2R4, respectively. In panel (a)
the contours of magnetic field (06:27 UT) are overploted (white:
positive, black: negative polarity, contour level: $\pm$100,
$\pm$500 G). The FOV of images is same as in Figure~\ref{preflare}.}
   \label{flare2}
   \end{figure}

\subsection{The M9.6 Class Flare} 
      \label{fl2}

\subsubsection{Chromospheric Evolution} 
\label{ce}

The M9.6 flare was preceded by a C3.8 event that also took place in
AR 10501 (see Figure~\ref{goes}). Due to a data gap in our H$\alpha$
observations between 07:21 UT and 07:34 UT, we could not locate the
brightenings related to this event. The H$\alpha$ evolution of the
M9.6 flare is shown in Figure~\ref{flare2}, where the different
ribbons are denoted as F2R1, F2R2, F2R3 and F2R4, respectively. The
first H$\alpha$ emission started at $\approx$ 07:35 UT in the
eastern section of F2R1 and propagated towards the western section
of the ribbon. The two central ribbons, F2R2 and F2R3, became
clearly visible and well-defined $\approx$ 2 minutes later. At that
time F2R4 became visible too. The two central ribbons clearly
separated with time, as typically seen in two-ribbon flares. The
emission in these ribbons was more intense and lasted longer than in
the outer ribbons (F2R1 and F2R4). The presence of a strong shear in
the field could be inferred from the locations of F2R2 and F2R3 at
both sides of the PIL; F2R2 was located significantly more to the
north-west than F2R3 (see Figures~\ref{ribbon_mdi2} and
\ref{flare2}), implying that the magnetic helicity was negative in
that region. This is consistent with the sign of the helicity
injected into the region by the clockwise rotation of N2$^\prime$
around P2 in the central bipole of the AR
(Section~\ref{mdi_evolution}; see also Figures~\ref{mdi} and
\ref{angle}).

In Figure~\ref{ribbon_mdi2}, we show the locations of the ribbons
for both flares. It can be seen that there are four ribbons in both
events and that they are located at almost the same positions, {\em
i.e.}, the flares were morphologically very similar. Ribbon F1R1
appears to be split in several bright parts in
Figure~\ref{ribbon_mdi2}, but it can be seen in Figure~\ref{flare1}
that it was in fact continuous. F2R2 and F2R3 are somewhat more
compact than F1R2 and F1R3, and F2R4 is shifted and rotated in the
clockwise direction compared to F1R4. A comparison of the
intensities reveals that, overall, the ribbons were brighter in the
M9.6 event.

Due to seeing limitations and the limited resolution of the
H$\alpha$ observations, it was sometimes difficult to clearly follow
the evolution of the flare ribbons. Still, we could observe an
interesting difference between the two events. Although the ribbons
were morphologically similar and located at almost the same places
in the AR, the temporal evolution of their emission was different:
during the first flare, the central ribbons, F1R2 and F1R3, appeared
before the outer ones (see Section~\ref{fl1}), while for the second
flare, the northern ribbons F2R1 and F2R2 appeared before the
southern ones.

The temporal evolution of the flare brightenings
and the northern ribbons appearence (see Section~\ref{Physical_Scenario}),
leads us to conclude that a so-called lateral breakout mechanism is the
most plausible scenario for the initiation of the second event.
Similar observational characteristics have been found in other events (see, {\em e.g.},
\opencite{Aulanier00}; \opencite{Gary04}; \opencite{Harra05}; \opencite{Williams05};
\opencite{Mandrini06}).

\subsubsection{Filament Eruption and Associated CME} 
      \label{filaments2}

After the first flare, the two elongated filaments with their
central parts, initially orientated along north-south direction,
change their orientation to an east-west direction (see
Section~\ref{filaments}). As time progresses, the filaments slowly
separate from each other. Around 07:35 UT, the northern filament
became partly unstable and its southern section moved quickly
towards the southern filament. The two filaments then interacted,
forming a structure with the shape of the letter ``Y''
(Figure~\ref{flare2}). The shape of this structure and its temporal
evolution indicate that the northern and the southern filaments were
merging (see the H$\alpha$ movie, ar10501-halpha.mpg). Moreover,
flare ribbons were developing during this evolution. Therefore, this
reconfiguration can be interpreted as being due to magnetic
reconnection between the magnetic configurations of the two
filaments.

Next, between 7:55 and 8:04 UT, the northern filament split in two
parts (see the H$\alpha$ movie). One part stayed mostly in place,
while the other part erupted. These two parts probably correspond to
the low and high sections of a partial filament eruption,
respectively \cite{Gibson06}. Flare ribbons were present, so
magnetic reconnection was involved in the eruption. The erupting
filament material could be seen moving south-west to a location in
between F2R2 and F2R3 and disappeared afterwards from the H$\alpha$
images. This eruption is interpreted as being due to the
reconnection of the magnetic arcade overlying the northern filament
with the arcade overlying the southern filament. This filament
eruption was probably associated with a full halo CME that became
visible in LASCO/C2 at 08:26 UT. This CME was faster ($\approx$ 670
km s$^{-1}$) than the one associated with the first flare.

The southern filament was still visible at 08:39 UT in H$\alpha$, {\it i.e.}, it was
probably not involved in the CME. However, it seems that later on it erupted as well
and did not reappear, at least as far as our H$\alpha$ data extend (see
Figure~\ref{flare2}d and the accompanying H$\alpha$ movie). Much later, after 11:30 UT,
the formation of a large-scale arcade was observed at the south of
the AR by EIT in 195 \AA . It was probably the signature of
reconnection occuring well behind the CME, involving the formation of closed loops
(due to relaxation of the magnetic field).

\section{The Magnetic Field Topology} 
      \label{topo}

\subsection{Magnetic Field Model} 
      \label{model}

The computation of the magnetic field connectivity and of its evolution
is relevant to understand the homologous characteristics of the two M-class
flares. Furthermore, the combination of the analysis of their
temporal and spatial evolution with the computation of the magnetic
field topology in the AR sets constrains on the energy release
mechanisms and the initiation processes of the studied events (see
Sections~\ref{fl1} and~\ref{fl2}).

The line of sight magnetic field of AR 10501 is
extrapolated to the corona using the discrete fast Fourier transform
method under the linear force-free field (LFFF) hypothesis
($\vec{\nabla} \times \vec{B}= \alpha \vec{B}$, where $\vec{B}$ is the magnetic field and
$\alpha$ is a constant related to the intensity of the coronal electric currents).
The method and its limitations are discussed in \inlinecite{Demoulin97}.

As boundary condition for the coronal magnetic model, we use the MDI magnetogram
closest in time to each flare onset (at 03:15 UT for the first flare and at 07:59 UT
for the second flare). For each magnetogram, we use a range of $\alpha$ values to
compute different magnetic field models.   For this particular study, the value
of the free parameter of each model, $\alpha$, is chosen such that the
ribbons observed during the flare main phases can be connected in
pairs. We have chosen this approach since no EUV or soft X-ray images,
in which coronal loops would have been clearly visible, were available
for times close to the studied events.

\inlinecite{Chandra10} analyzed the magnetic field connectivity and
computed the magnetic helicity injection in AR 10501 in search for
an explanation of the link between the flares and CMEs, that
occurred on 18 November 2003, and the interplanetary magnetic cloud
observed at 1 AU on 20 November. The aim of that study was to
understand why an AR that had a global negative magnetic helicity
could expel a positive helicity magnetic cloud to the interplanetary
medium. These authors found that a single $\alpha$ value was not
sufficient to represent the connectivity between different
polarities in the AR. The observed loops in some regions of the AR
could be modeled using a negative $\alpha$ value, while a positive
$\alpha$ value was needed for others. The computed maps of the
magnetic helicity flux density also showed mixed helicity signs.
Furthermore, there was observational evidence of different
chiralities in different sections of the filaments seen in
H$\alpha$. It was finally concluded that positive magnetic helicity
had been injected into a connectivity domain at the south of the AR.
The CME and associated magnetic cloud were ejected from this region.

The mixed helicity sign pattern just described seems to be present
also in the AR configuration on 20 November. Different sections of
the observed H$\alpha$ ribbons can be connected in pairs using
$\alpha$ values ranging from negative to positive values. In all
cases, the value of $\alpha$ is low, between $-9.4
\times$10$^{-3}$Mm$^{-1}$ and $7.5 \times$ 10$^{-3}$Mm$^{-1}$. We
will discuss the magnetic field connectivity and its implications
for the flare initiation process in Section~\ref{QSL}, after briefly
introducing the definition of QSLs and the method used to compute
their locations.

\begin{figure}
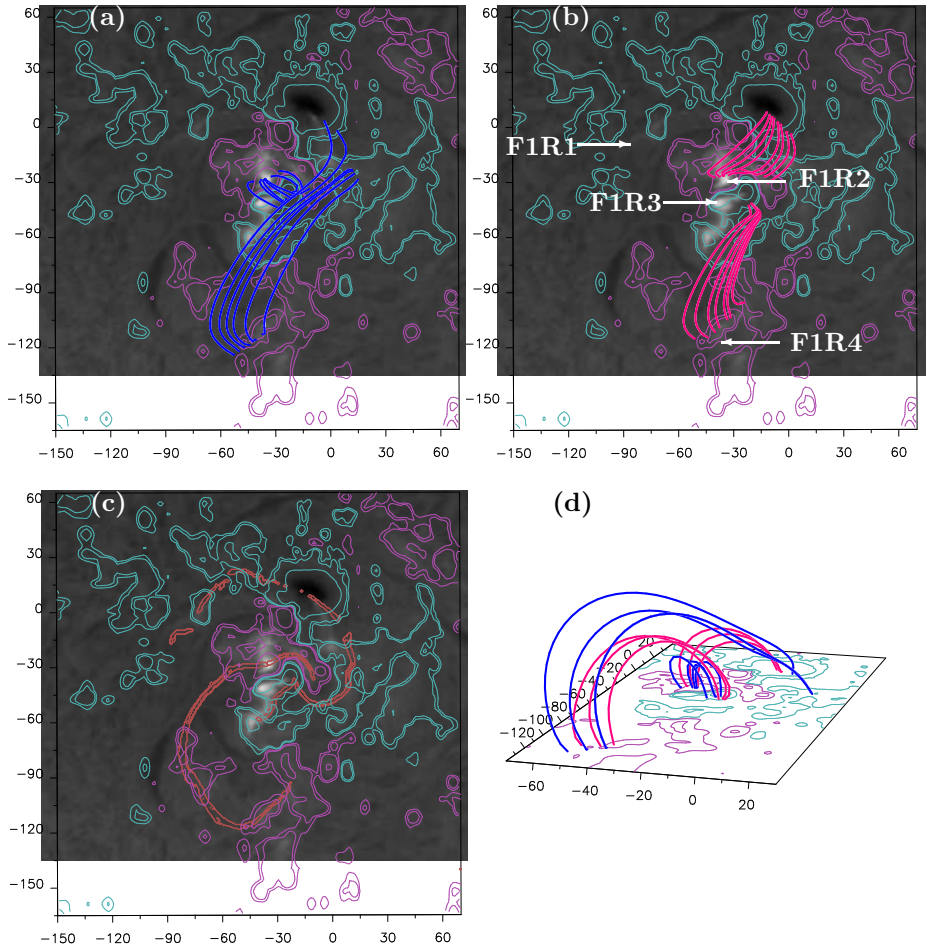
 
\vspace*{-0.01\textwidth}
\centerline{\hspace*{-0.01\textwidth}
\includegraphics[width=0.50\textwidth,viewport=25 100 500 600,clip=]{fig9a}
\hspace*{-0.02\textwidth}
\includegraphics[width=0.50\textwidth,viewport=25 100 500 600,clip=]{fig9b}
\color{white} \bf\put(-320,175){(a)}
\color{white} \bf\put(-145,175){(b)}
$ \color{white} \put(-135,130){\vector(1,0){20}} \color{white} \put(-162,126){F1R1}$
$ \color{white} \put(-60,116){\vector(-1,0){22}} \color{white} \put(-55,114){F1R2}$
$ \color{white} \put(-110,108){\vector(1,0){20}} \color{white} \put(-138,105){F1R3}$
$ \color{white} \put(-70,55){\vector(-1,0){22}} \color{white} \put(-66,52){F1R4}$
}
\centerline{\hspace*{-0.05\textwidth}
\includegraphics[width=0.50\textwidth,viewport=25 100 500 600,clip=]{fig9c}
\hspace*{-0.02\textwidth}
\includegraphics[width=0.50\textwidth,viewport=25 100 500 600,clip=]{fig9d}
\color{white} \bf\put(-320,175){(c)} \color{black}
\bf\put(-145,175){(d)} } \caption{ (a,b): H$\alpha$ image of the
first flare at 02:17 UT overlaid by magnetic field lines computed
from either side of the photospheric traces of the QSLs shown in
(c). (c): Intersection of the QSLs with the photospheric plane
computed from the MDI magnetogram at 03:15 UT, overlaid on the
H$\alpha$ image. Panels (a-c) are drawn in the observer's point of
view. (d): Perspective view of the field lines shown in panels (a)
and (b). The field lines have been selected to illustrate the
coronal linkage at the edges of QSLs. The field of view is a portion
of the one shown in (a-c). Field lines drawn in blue (red) represent
loops before (after) reconnection at a QSL, as inferred from the
analysis of the flare evolution (see
Section~\ref{Physical_Scenario}). All panels correspond to the
magnetic field model with the lowest (negative) $\alpha$ value. The
negative and positive magnetic field contour levels ($\pm$ 50, 100,
500 gauss (G) in (a--c), and $\pm$ 100, 500 G in (d)) are represented in cyan
and pink colours, respectively. }
   \label{extra1}
   \end{figure}

\begin{figure}
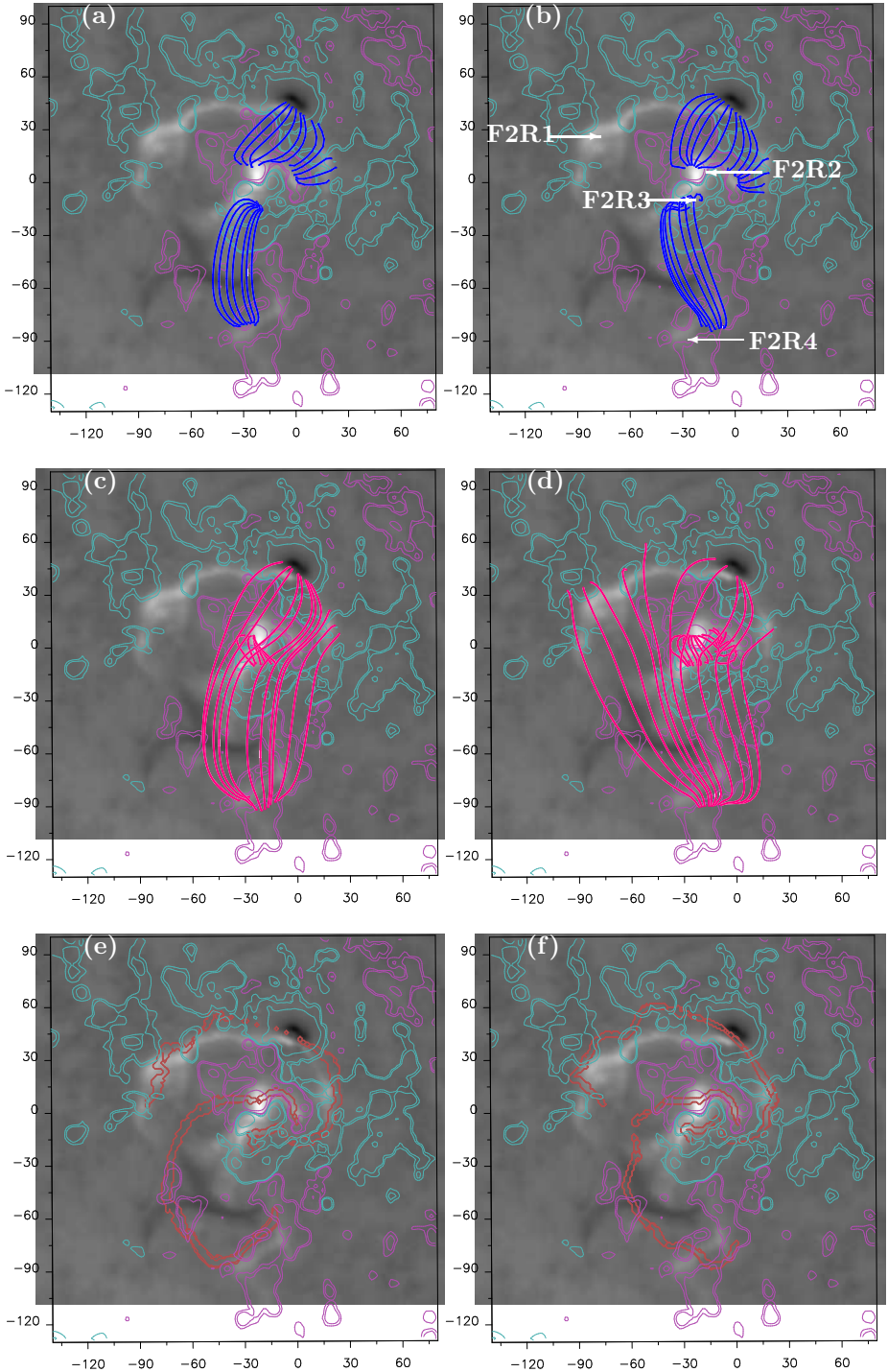
 
\centerline{\hspace*{-0.01\textwidth}
\includegraphics[width=0.50\textwidth,viewport=25 100 500 600,clip=]{fig10a}
\hspace*{-0.02\textwidth}
\includegraphics[width=0.50\textwidth,viewport=25 100 500 600,clip=]{fig10b}
\color{white} \bf\put(-320,175){(a)}
\color{white} \bf\put(-145,175){(b)}
$ \color{white} \put(-135,130){\vector(1,0){20}} \color{white} \put(-160,128){F2R1}$
$ \color{white} \put(-55,116){\vector(-1,0){22}} \color{white} \put(-51,114){F2R2}$
$ \color{white} \put(-105,105){\vector(1,0){20}} \color{white} \put(-130,102){F2R3}$
$ \color{white} \put(-70,50){\vector(-1,0){22}} \color{white} \put(-68,46){F2R4}$
}
\centerline{\hspace*{-0.05\textwidth}
\includegraphics[width=0.50\textwidth,viewport=25 100 500 600,clip=]{fig10c}
\hspace*{-0.02\textwidth}
\includegraphics[width=0.50\textwidth,viewport=25 100 500 600,clip=]{fig10d}
\color{white} \bf\put(-320,175){(c)}
\color{white} \bf\put(-145,175){(d)}
}
\centerline{\hspace*{-0.05\textwidth}
\includegraphics[width=0.50\textwidth,viewport=25 100 500 600,clip=]{fig10e}
\hspace*{-0.02\textwidth}
\includegraphics[width=0.50\textwidth,viewport=25 100 500 600,clip=]{fig10f}
\color{white} \bf\put(-320,175){(e)} \color{white}
\bf\put(-145,175){(f)} } \caption{ (a-d): H$\alpha$ image of the
second flare at 07:39 UT overlaid by magnetic field lines computed
from either side, or close to, the photospheric traces of QSLs shown
in (e) and (f). (e,f): Intersection of the QSLs with the
photospheric plane, computed from the MDI magnetogram at 07:59 UT
overlaid on the H$\alpha$ image at 07:39 UT. The panels in the left
column correspond to the magnetic field model using the lowest
(negative) $\alpha$ value (with some field lines computed from a
potential field model added), the panels in the right column
correspond to the model using the highest (positive) value for
$\alpha$. Field lines drawn with continuous blue (red) lines
represent loops before (after) reconnection at a QSL, as inferred
from the analysis of the flare evolution (see
Section~\ref{Physical_Scenario}). The negative and positive magnetic
field contour levels ($\pm$ 50, 100, 500 G) are shown in cyan and
pink colours, respectively. All panels have been drawn in the
observer's point of view.}
   \label{extra2}
   \end{figure}

\subsection{QSLs Locations and Flare Ribbons}
\label{QSL}

QSLs are defined as regions where there is a drastic change in field
line connectivity, as opposed to the extreme case of separatrices
where the connectivity is discontinuous (see references in
Section~\ref{S-Introduction}). The location of QSLs can be
determined by computing the squashing degree $Q$ \cite{Titov02}. A QSL
is defined where $Q \gg 2$ (the value $Q=2$ is the lowest possible
one). The squashing degree increases where connectivity gradients
grow, and it becomes infinitely large where the field line mapping
becomes discontinuous, {\it i.e.}, for separatrices. By definition,
$Q$ is uniquely defined along a field line by $(\vec{B} \cdot
\vec{\nabla}) Q = 0$.

The numerical procedure used to determine the values of $Q$ has been
thoroughly discussed by \inlinecite{Aulanier05}. The magnetic field
model used here takes observed magnetograms as the boundary
condition. Therefore, the presence of parasitic polarities results
in multiple QSLs. However, we only show those corresponding to the
highest values of $Q$ ($\log_{10}Q$ above $\approx 6$). In other
words, only the thinnest QSLs, those where thin electric current
sheets favourable for reconnection are most likely to build up, are
shown in Figures~\ref{extra1} and \ref{extra2}. It is worth noting
that, in all our computations of the coronal magnetic field, no
magnetic null points were found that could be related to the large
scale connections between the flare ribbons, i.e. there are no
separatrices which can explain the location of the observed
H$\alpha$ ribbons.

Figures~\ref{extra1} and~\ref{extra2} show the intersections of the
QSLs with the photospheric plane, together with the locations of
H$\alpha$ ribbons at times corresponding to the main phases of the
M1.4 and M9.6 flares, respectively. The linkage of the coronal
magnetic field is outlined by field lines having footpoints on
either side, and close to, the photospheric traces of the QSLs.

Two elongated QSLs are present in the AR configuration, as has been
found in cases analyzed previously (\opencite{Demoulin97};
\opencite{Mandrini97}). The flare ribbons for both events are
located at, or close to, the photospheric QSL traces. For the first
flare, we only show the results for a magnetic field model with
$\alpha$ = -9.4 $\times$10$^{-3}$Mm$^{-1}$ (Figure~\ref{extra1}),
while for the second flare we show the results for both values of
$\alpha$, -9.4 $\times$10$^{-3}$Mm$^{-1}$ and 7.5 $\times$
10$^{-3}$Mm$^{-1}$ (Figure~\ref{extra2}).
The different sets of field lines drawn in these figures illustrate the main
connectivity domains in AR 10501.
We have also added some field lines computed using a potential field model
in Figures~\ref{extra2}a and c, since the linkage between the ribbons
is better represented by them along some sections of the QSLs. In
what follows we discuss the results for the M9.6 flare. The same
results are valid for the M1.4 event.

For the second flare, one set of field lines connects flare ribbons
F2R1 to F2R2, a second set F2R2 to F2R3, a third set F2R3 to F2R4,
and a fourth set F2R1 to F2R4 (see Figure~\ref{extra2}).  As
discussed in Section~\ref{model}, our choice of $\alpha$ was done
such that the flare ribbons could be connected by field lines in
pairs. Using either a negative or positive $\alpha$ value, ribbons
F2R2 and F2R3 overlap very well with photospheric QSL traces along
all their extensions (Figures~\ref{extra2}). The same is true for
the western part of F2R1. Its eastern part, however, does not
overlap that well with the QSL trace, most probably because the QSL
is broad there (lower $Q$ values). Therefore, its location is more
sensible to the coronal current distribution. Finally, ribbon F2R4
is located closer to a QSL trace in the model with negative
$\alpha$.

A linear force-free model implies that $\alpha$ is constant in
the whole AR. However, from our previous discussion (Section~\ref{model}),
we can assume that this is not the case for AR 10501, since
it seemed to retain the same mixed helicity pattern found on 18 November \cite{Chandra10}.
Therefore, the partial lack of coincidence between the spatial locations of
QSLs and flare ribbons comes most probably from the intrinsic limitations
of our modelling. In spite of this, there is a global agreement
between the QSL traces and flare ribbons for both events
(Figures~\ref{extra1} and~\ref{extra2}). Furthermore, these traces
are similar even though QSLs have been computed using two different
magnetograms as boundary conditions (at 03:15UT and at 07:59 UT);
this means that the magnetic configuration involved in both flares
remained approximately the same, explaining the homology between them.

In addition, we found that for all the $\alpha$ values used, the
resulting overall pattern of the QSL traces was similar. This
indicates that the locations of QSLs are mainly determined by the
photospheric distribution of the magnetic polarities. For both
events, the flare ribbons can be connected in pairs by field lines
whose foopoints lie on both sides of a QSL (Figures~\ref{extra1}
and~\ref{extra2}). We therefore conclude that magnetic reconnection
at QSLs is at the origin of both homologous flares.

\subsection{Physical Scenario of the Eruptive Flares} 
      \label{Physical_Scenario}

The comparison of the observed flare ribbons and the photospheric
trace of QSLs shows that reconnection at QSLs is involved in the
flares, but it does not explain which are the reconnected field lines,
since the spatial comparison is not precise enough to tell on which
side of a QSL is a given ribbon, {\em i.e.}, if reconnected field lines are
connecting the inner (resp. the outer) pair of ribbons or if they
are connecting the northern (resp. southern) pair of ribbons.
However, there are several observational clues to determine the connectivities
after reconnection. A classical one is to observe flare loops in UV
or/and in H$\alpha$. Unfortunately, in the analyzed observations,
there is no clear evidence of flare loops. A related clue is the observation
of hard X-rays sources at the footpoints or/and near the top of
reconnected loops. Those are observed during the first flare
(Figure~\ref{rhessi}b). Another clue is the increasing separation
with time of a pair of ribbons located at both sides of a PIL, as
observed in the second flare.

The main evidence for the field connections after reconnection in
the M1.4 flare are the spatial locations of the three hard X-rays
sources observed by RHESSI (Figure~\ref{rhessi}b). They imply that
reconnected field lines connect F1R1 to F1R2 and, by inference, also
F1R3 to F1R4 (drawn with red lines in Figures~\ref{extra1}b and d).
There are no detectable hard X-ray sources, and the H$\alpha$
ribbons are weaker, for these southern connections, most probably
because the magnetic configuration is very asymmetric in magnetic
field strength, so that there is only a low energy input into the
southern reconnected loops. The new connections (F1R1 to F1R2, F1R3
to F1R4) are each located above an H$\alpha$ filament. It can be
assumed that the downward magnetic field tension of these new
connections increases the stability of the magnetic configurations
of the filaments. Indeed, these two filaments were staying
unperturbed during and after the M1.4 flare. Finally, we infer from
the above observational facts that the connections before
reconnection are field lines connecting F1R2 to F1R3 and F1R1 to
F1R4 (drawn with blue lines in Figures~\ref{extra1}a and d). This
conclusion is coherent with the strong shearing motion observed in
between F1R2 and F1R3; such a motion is likely to store magnetic
energy and to expand short connections, forcing them to reconnect
with the overlying large connections, as in the MB model. Recalling
that we have also found evidence of the TC model in the pre-flare
stage (Figure~\ref{preflare}), we conclude that the physical
mechanism at work during the eruption is first TC, followed by an
MB-like mechanism.

While the flare ribbons have nearly the same locations in the M9.6
flare as in the previous flare, we found evidence that reconnection
proceeded in the opposite way, as follows. For the M9.6 flare, the
central ribbons clearly separated with time, indicating that new
magnetic connections were formed between F2R2 and F2R3 (see the
H$\alpha$ movie in the interval 7:40-8:00 UT). Moreover, between
7:55 and 8:04 UT, a part of the northern filament erupted (see
Section~\ref{filaments2}). We interprete this eruption as being due
to the reconnection of the magnetic arcade overlying this part of
the filament ({\em i.e.}, the one between F2R1 and F2R2) with the
southern arcade (the one between F2R3 and F2R4; see
Figure~\ref{extra2}). Below the southern arcade, the filament was
also destabilized, but later on (after 8:10 UT). We conclude that,
in the M9.6 flare, the overlying arcade connecting F2R1 to F2R2
reconnected with the arcade connecting F2R3 to F2R4, {\em i.e.}, in
the reverse direction than in the MB model (see
Section~\ref{S-Introduction}). This is sometimes referred to as
``lateral breakout'' (see the references in Sect.~\ref{ce}).

Although the flares are homologous, we conclude that reconnection
proceeded in opposite directions for the M1.4 and M9.6 flares.
Reconnection occurring successively in a reverse direction during a
a flare has been also reported by \inlinecite{Goff07} and by
\inlinecite{Karpen98} as result of an MHD simulation. However, in
the flare studied by \inlinecite{Goff07}, the reconnection took
place in successive phases of the same flare and the reversal of the
reconnection direction in the last phase was interpreted as a
``bouncing back'' due to a too strong forcing in the previous
eruptive phase. In the homologous flares studied here, the reversal
has another origin, as we discuss below.

First, the strong shearing induced by the motion of N2$^\prime$ (and
by other smaller negative polarities) leads to the expansion of the
field lines connecting P2 to N2$^\prime$. There is also diffusion of
the magnetic field towards the PIL, which is accompanied by flux
cancelation and, most probably, by the build up of a flux rope, as
in the MHD simulation of \inlinecite{Aulanier10}. The pre-flare
brightenings (Figure~\ref{preflare}) are a trace of this slow
reconnection process.  Then, at some point of the evolution, either
a fast quadrupolar reconnection process starts or the flux rope
becomes unstable and the M1.4 flare occurs. Since the motion of
N2$^\prime$ continues (Figure~\ref{angle}), we could expect that a
similar flare, with the same reconnection direction, would occur
later on. However the build-up of magnetic stress is occurring also
along other PILs of the AR, since there is a large photospheric
dispersion of the magnetic field from the main spot (see the MDI
movie). As described above, converging flows are increasing the
magnetic shear, force flux cancelation, and probably also build a
flux rope. This process is especially active in between P1, P2, and
the eastern diffuse negative polarity (both large flows and a large
amount of magnetic flux are involved). We suggest that this process
was strong enough to build up a too stressed, hence unstable, field
configuration at the location of the northern filament, ({\em i.e.,}
between F2R1 and F2R2), inducing the M9.6 flare. Therefore, since
the eruption starts in a connectivity domain different from that of
the first flare, the successive reconnections proceed in the
opposite order in the two flares.

\section{Conclusions} 
      \label{conclusion}

We studied two homologous flares on 20 November 2003 that occurred
in AR 10501. The active region is characterized by multiple
polarities and by a significant rotation ($\approx 110^{\circ}$) of
a negative polarity around the corresponding positive polarity of an
emerging bipole. This continuous rotation supplies free energy to
the active region. The AR is also characterized by a large diffusion
of the magnetic flux from a main sunspot (polarity N1), accompanied
by the convergence and subsequent cancellation of opposite
polarities at the photospheric inversion lines (PILs). This process
also increases the stress of the coronal magnetic field.

Four flare ribbons are observed in each flare, indicating that
reconnection occurs in a quadrupolar configuration. The photospheric
magnetic field was extrapolated, using the linear, or constant
$\alpha$, force-free-field approximation. Due to the
complexity of the AR, in particular, evidence of
magnetic helicity with mixed signs two days before, we computed the
QSL locations using positive as well as negative $\alpha$ values. In
agreement with previous studies (see references in
Section~\ref{S-Introduction}), the H$\alpha$ flare ribbons are found
in the vicinity of the photospheric trace of QSLs. They are
magnetically  linked in pairs, as expected for a quadrupolar configuration.
Since the large-scale connections of the coronal field are only
weakly sheared, the photospheric traces of QSLs are located at almost
the same positions for cases with $\alpha <0$ and $>0$. Thus, the
QSL locations are mainly determined by the spatial distribution of
the photospheric polarities, and to a much lesser extent by the coronal
currents.  However, we found that $\alpha$ values of different sign are required
to model the magnetic connection between different pairs of ribbons, which
supports the presence of mixed helicity in the AR.

Thanks to the high time cadence of the H$\alpha$ observations, we
also observed the fast drift of the emission along one ribbon
(F2R1). This can be interpreted as a slip-running magnetic
reconnection process \cite{Aulanier07}, which is a characteristic of
reconnection at QSLs with finite thickness (in contrast to an
instantaneous change of connectivity for reconnection across
separatrices).

The location of the pre-flare brightenings suggests that the first
flare was initiated by ``tether cutting'' at the central PIL. A
significant motion of a nearby magnetic polarity was observed for
about 8 hours before the flare, providing a very sheared core field,
in agreement with vector magnetogram observations. Later on, this
core field erupted, accompanied by the formation of a four-ribbon
flare. Tether cutting reconnection at the PIL, {\em i.e.,} the
transformation of a sheared arcade into a flux rope, appears to be
the mechanism starting the eruption.  Subsequent quadrupolar
reconnection is evident from the observed pattern of the flare
ribbons; this presumably destabilized the configuration even more
(as in the magnetic breakout model).  The flux rope eruption itself
may have been driven by an instability or catastrophic loss of
equilibrium. Present observations are not sufficient to separate the
respective roles of these mechanisms.

The second flare had a very similar spatial organization of the
H$\alpha$ ribbons, {\em i.e.}, it was homologous to the first one.
However, we found observational evidence that magnetic reconnection
occurs in a direction reverse to that of the first flare ({\em
e.g.}, the pieces of evidence are the evolution of the ribbon
separation from the central PIL and the locations where the
filaments erupt, below the lateral field connections of the
quadrupolar configuration). This is surprising at first, since the
observed shearing motions continued well after the time of the
flares. However, there was also a significant diffusion of the
magnetic field around the main sunspot, especially in the northern
part of the AR. This leads to converging motions and magnetic
cancelation along the nearby PIL, as well as to an increase of the
magnetic shear. We conclude that the second flare occurred as this
process destabilized the northern lateral connectivity domain of the
quadrupolar configuration, {\em i.e.}, the second eruption started
in a different domain than the first one. This resulted in a flare
with similar ribbons, but with a reconnection direction opposite to
the one during the first flare.

More generally, we suggest that in a complex magnetic configuration
with several connectivity domains (separated by separatrices or
QSLs), horizontal motions of the polarities, magnetic diffusion
toward the PILs, as well as newly emerging flux, will all compete in
increasing the local stress of each connectivity region. Depending
on the relative importance of these processes, but also on the
previous history (like the launch of a CME from one of the
connectivity domain), the stressed magnetic field of one of the
connectivity domains will become the next unstable region. As long
as the photospheric flux distribution is not drastically changed,
the successive flares will be homologous, {\em i.e.}, they will
display the same global pattern of flare ribbons. This is because,
for multipolar fields, QSL locations are determined mainly by the
global organization of the photospheric polarities and QSLs are
structurally stable (while separatrices can appear or disappear
after a bifurcation of the topology). In the absence of significant
emergence, the only possible change expected from one flare to the
next is the location of the next unstable connectivity domain, {\em
i.e.}, the location where the next unstable pre-CME core field is
forming.

\begin{acks}
R.C. thanks Centre Franco-Indien pour la Promotion de la Recherche
Avanc\'ee (CEFIPRA) for his postdoctoral grant. C.H.M. thanks the
Argentinean grants: UBACyT X127 and PICT 2007-1790 (ANPCyT). C.H.M.
is a member of the Carrera del Investigador Cient\'{i}fico, CONICET.
The research leading to these results has received funding from the
European Commission's Seventh Framework Programme (FP7/2007-2013)
under the grant agreement n$^o$ 218816 (SOTERIA project,
www.soteria-space.eu). Financial support by the European Comission
through the SOLAIRE network (MTRM-CT-2006-035484) is also gratefully
acknowledged. We acknowledge the use of TRACE data. MDI data are a
courtesy of SOHO/MDI consortium. SOHO is a project of international
cooperation between ESA and NASA. B.S, C.H.M, RC have started and
discussed this work in the frame of the ISSI workshop chaired by Dr.
Consuelo Cid (2008-2010). The authors acknowledge financial support
from ECOS-Sud (France)/MINCyT (Argentina) through their cooperative
science program (N$^o$ A08U01). We also thank the anonymous referee's constructive comments and suggestions.
\end{acks}



\mbox{}~\\
\bibliographystyle{spr-mp-sola}
\IfFileExists{\jobname.bbl}{} {\typeout{}
\typeout{***************************************************************}
\typeout{***************************************************************}
\typeout{** Please run "bibtex \jobname" to obtain the bibliography}
\typeout{** and re-run "latex \jobname" twice to fix references}
\typeout{***************************************************************}
\typeout{***************************************************************}
\typeout{}}

\end{article}
\end{document}